%Paper: gr-qc/9307008
%From: sakellar@UNIV-TOURS.FR
%Date: Fri, 09 Jul 1993 15:27:31 EDT

\magnification=1200
\baselineskip=18pt
\centerline{{\bf Gravitational Effects on Domain Walls with Curvature
Corrections}}
\vskip 1 true cm
\centerline{C. Barrab\`es$^{\dagger, \ddagger, \diamond}$,
B. Boisseau$^{\dagger, \triangleleft}$ and M. Sakellariadou$^{\dagger, \star}$}
\vskip 1 true cm
\centerline{$^{\dagger}$Laboratoire de Mod\`eles de Physique Math\'ematique,}
\centerline{Facult\'e de Sciences et Techniques,}
\centerline{Parc de Grandmont, 37200 Tours, France.}
\vskip 0.5 true cm
\centerline{$^{\ddagger}$
D\'epartement d'Astrophysique Relativiste et Cosmologie,}
\centerline{Observatoire de Paris, 92190 Meudon, France.}
\vskip 1 true cm
\centerline{ Abstract}
\vskip 1 true cm
We derive the effective action for a domain wall with small thickness in curved
spacetime and show that, apart from the Nambu term, it includes a
contribution proportional to the induced curvature. We then use this action
to study the dynamics of a spherical thick bubble of false vacuum (de Sitter)
surrounded by an infinite region of true vacuum (Schwarzschild).
\vskip 1 true cm
PACS 98.80Dr; 04.20.Jb
\vfill\eject
\vskip 1 true cm
\centerline{I. Introduction}
\vskip 1 true cm
An intriguing implication of unified gauge theories is the possibility,
within certain models, of the coexistence of two phases separated by
a wall, which, at first approximation, can be seen as an infinitely thin
bubble whose history is a timelike hypersurface$^1$.
Within this context, there is a number of articles which appeared in the
literature studying the dynamics of bubbles with a surface layer described
by the Nambu action of a domain wall. Such an action can be obtained from a
field theory, by considering at first approximation, that the field is
condensed along a three-dimensional timelike hypersurface, whose area gives
the effective Nambu action. This last approach yields a description of the
dynamics of a domain wall under the assumption that its dimensions are
much greater than its thickness.

During the last few years, a number of authors studied the finite thickness
corrections to the Nambu action. Gregory and Gregory {\it et al}$^2$ have
calculated the leading-order corrections to the equation of motion
for a finite thickness domain wall, due to its extrinsic curvature and its
self-gravity. Their conclusion was that the effective wall action must
include, apart from the Nambu term, a contribution proportional to the
induced curvature. On the other hand, Silveira and Maia$^3$ expanding the
effective action in powers of the thickness, concluded that, in flat
spacetime, there is a first order correction term  on the
mean curvature and two second order correction terms: one  depending
on the induced curvature and the other one depending on the Gaussian
curvature.
While we were writing our work,  a preprint
of Larsen$^4$, regarding the finite thickness corrections to the Nambu
action for a curved domain wall in Minkowski spacetime, came to our
attention.
Following the same approach as Silveira and Maia$^3$, Larsen$^4$ found that
the correction term is proportional to the Ricci curvature of the induced
metric.
Hence, since the results in the literature seem to disagree, the first aim
of the study we present in this article is to obtain a consistent
derivation of the effective action for a curved thick domain wall
in  curved spacetime. Our result agrees with that of Gregory$^2$,
Larsen$^4$ and also Letelier$^5$, who studied the dynamics of test bubbles
with curvature corrections in  flat spacetime. The second aim of our work
is to examine the effect of the curvature corrections on the bubble
dynamics including gravitational effects.
An analogous study   was done by Blau {\it et al}$^6$, who studied
the dynamics of a spherically symmetric region of false vacuum
separated by a Nambu domain wall from an infinite region of true
vacuum.

This paper is organized as follows: In section II we derive the
effective action and the energy momentum tensor for a domain wall
with finite thickness corrections. In section III we study the
dynamics of such  domain wall separating two spherically symmetric
static
geometries. We then focus on the particular case of a spherical
 thick bubble of false vacuum (de Sitter)
surrounded by an infinite region of true vacuum (Schwarzschild).
We finally compare our results with those of Blau {\it et al}$^6$.

Our system of units is such that $\hbar = c = 1$.

\vskip 1 true cm
\centerline{II. Domain wall with curvature corrections.}
\vskip 1 true cm
In this section we will derive the effective action and effective
energy momentum tensor for a test domain wall moving in empty space,
when curvature corrections
are taken into account. We consider a real scalar field $\phi $ with
action
$$A[\phi , g]\ = \ \int \sqrt{-g}\  {\cal L}  (\phi , g)\
 d^4 x\ ,\eqno(II. 1)$$
in a spacetime manifold $\cal M $ with signature $(-+++)$. The Lagrangian
of $\phi $ reads
$$ {\cal L} (\phi , g)\ = \ -{1 \over 2} g^{\mu \nu}\ \partial _{\mu}\phi \
\partial _{\nu}\phi \ - {1 \over 2}\
 \lambda (\phi ^2 -\eta ^2)^2, \ \ \lambda >0\ .
\eqno(II.2)$$
Taking the extremum of the action we obtain the field equation
$$(1/\sqrt {-g}) \ \partial _{\mu}\
(\sqrt{-g} g^{\mu \nu}\partial_\nu \phi)\ -\
2\lambda \phi \ (\phi^2 -\eta ^2)\ =\ 0\ .\eqno(II.3)$$
A solution of the above field equation describes a membrane-like structure
provided there exists a timelike hypersurface $\Sigma $, such that the field
$\phi$ vanishes on $\Sigma $, while at sufficiently large distance from it,
$\phi $ takes the value $+\eta $ on one side and $-\eta $ on the
other one. Such a solution is called  domain wall and  can be considered
as a macroscopic membrane of a certain width, whose average history is
given by $\Sigma $. In flat spacetime the explicit form of the solution
of Eq.(II.3) has been obtained$^7$ in the particular case corresponding
to a planar static domain wall. This particular solution, representing a
flat membrane of width $\delta $ lying along the plane $z=0$, reads
$$\phi_0(z)\ =\ \eta \ \tanh(z/\delta)\ ,\eqno(II.4)$$
where
$$\delta \ = \ \eta^{-1}\ \lambda ^{-1/2}\ .\eqno(II.5)$$
The action corresponding to $\phi_0$ is
$$A_0\ =\ -(\eta /\delta)^2\ \int \cosh^{-4}(z/\delta)\ d^4x\ =\ -\mu
\int \ dt \ dx \ dy\ ,\eqno(II.6)$$
where the mass per unit area $\mu$ is given by
$$\mu\ =\ (2\eta^2/\delta)\ \int_0^{\infty} dz \ \cosh^{-4}z\ \approx \
2 \eta^2/ \delta\ .\eqno(II.7)$$
Here, we are interested in the effective action of a general
solution of the field equation (II.3), describing membrane-like structures.

We find it convenient to introduce in a neighborhood of $\Sigma$
 Gaussian coordinates $x^{\alpha}=(\tau ^
A ,\rho)$, where $\tau ^A \ (A= 0,1,2)$ are the parameters on $\Sigma $ and
$\rho $ is the proper length along the geodesics orthogonal to
$\Sigma $. In this system, the metric takes the form:
$$ {\cal G}_{\alpha \beta} \ =\  ({\cal G}_{A B}(\tau, \rho),\   {\cal G}_
{\rho \rho}=1,\  {\cal G}_{A \rho}=0\ )\  .\eqno(II.8)$$
Let $\partial_{\alpha } = (\partial_A,\partial_{\rho})$ be the holonomic
basis. We can then associate with each point on $\Sigma $, four linearly
independent vectors $(e_A,N)$ defined as
$$e_A\ =\ (\partial_A)_0\ ,\ N\ =\ (\partial _{\rho})_0\ .\eqno(II.9)$$
The induced metric
on $\Sigma $ and its extrinsic curvature are, respectively,
$$\gamma _{A B}\ = \ e_A \  \cdot e_B \ ,\eqno(II.10)$$
$$K_{A B}\ =\ \nabla _A\  N \ \cdot e_B \ =\ -N\ \cdot \
   \nabla _A  \ e_B\ .\eqno(II.11)$$
A second order series expansion of the metric in the neighborhood of
$\Sigma $ gives
$${\cal G} _{A B}\ = \ \gamma _{A B} + 2\rho K_{A B} + \rho^2 [K_A^{\ \ C}
K_{B C}-(R_{A \rho B \rho})_0] +{\cal O}(\rho^3)\ ,\eqno(II.12)$$
where $R_{\alpha \beta \gamma \delta}$ stands for the Riemann
tensor. The inverse metric tensor ${\cal G} ^{\alpha \beta}$ is
$$ {\cal G}^{A B}\ = \ \gamma^{A B}- 2\rho K^{A B} + \rho^2 \  [3K^{A C}
K_C^{\ \ B}+(R^{A \ \ \ B}_{\ \  \rho \ \ \ \ \rho})_0]
+ {\cal O}(\rho^3)\ ,
\eqno(II.13)$$
where $\gamma^{A B}$ is the inverse of the induced metric $\gamma _{A B}$;
$${\cal G}^{A \rho}\ =\ {\cal G}_{A \rho}\ =\ 0\ ; \eqno(II.14)$$
and
$${\cal G}^{\rho \rho}\ = \ {\cal G}_{\rho \rho}\ =\ 1\ .\eqno(II.15)$$
As it can be easily verified
$$\sqrt{-{\cal G}}\ =\ \sqrt {-\gamma}[1+\rho K + {1 \over 2}
\rho ^2 (K^2 -K^{A B}
K_{A B}-(R_{\rho\rho})_0)+{\cal O}(\rho^3)]\ ,\eqno(II.16)$$
where $\gamma = det (\gamma_{A B}), K=Tr(K^A_{\ \ B})$ and $R_{\rho \rho}$
is the $(\rho \rho)$-component of the Ricci tensor. Applying the
Gauss-Codazzi equation
$$-2G_{\alpha \beta} N^{\alpha} N^{\beta}\ =\ ^{(3)}R + K_{A B}K^{A B}-K^2\ ,
\eqno(II.17)$$
where $^{(3)}R$ denotes the induced Riemannian curvature on $\Sigma $ and
$G_{\alpha \beta}$ stands for the Einstein tensor, we
obtain that in an empty curved spacetime
Eq.(II.16) reduces to
$$\sqrt{-{\cal G}}\ =\ \sqrt {-\gamma}\ [1+\rho K + {1 \over 2}
\rho^2\  ^{(3)}R +
{\cal O}(\rho^3)]\ .\eqno(II.18)$$

Let us now concentrate on solutions close to the planar static
one. To do so, we suppose that the curvature radius of the membrane is large
enough to assume that the solution $\phi$ is near to $\phi_0(\rho)$, where
now $\rho $ replaces $z$. Thus,
$$\phi \ = \ \phi_0(\rho)+\phi_1(x^{\alpha})\ ,\eqno(II.19)$$
where $\phi_1$ denotes the perturbative term, which is such that $\phi_1\ \ =0$
on $\Sigma $ and $\phi_1$ goes to zero far from the hypersurface $\Sigma$.
Expanding the Lagrangian
${\cal L}$ to second order in $\phi_1$, one gets
$${\cal L}\ =\ L_0 +L_1+L_2\ ,\eqno(II.20)$$
where
$$L_0\ =\ -{1 \over 2} \partial _{\rho}\phi_0 \partial _{\rho}\phi_0
-{1 \over 2}\lambda (\phi_0^2-\eta^2)^2\ ;\eqno(II.21)$$
$$L_1\ =\ -\partial _{\rho}\phi_0 \partial _{\rho}\phi_1 - 2\lambda \phi_0
(\phi_0^2-\eta^2)\phi_1\ ;\eqno(II.22)$$
$$L_2\ =\ -{1 \over 2}{\cal G}^{\alpha \beta}\partial_{\alpha }\phi_1 \partial
_{\beta}\phi_1-\lambda(3\phi_0^2-\eta^2)\phi_1^2\ .\eqno(II.23)$$
Note that the zero order term $L_0$ is identical to the Lagrangian
for the planar static solution $\phi_0$. Using the field equation
 for the planar solution $\phi_0$, {\it i.e.,}
$$\partial_{\rho \rho}\phi_0(\rho)-2\lambda\phi_0(\phi_0^2-\eta^2)\ = \ 0\ ,
\eqno(II.24)$$
the first order term $L_1$ reduces to
$$L_1\ =\  -\partial_{\rho}(\phi_1 \partial_{\rho}\phi_0)\ .\eqno(II.25)$$
Moreover, for the solution given by Eq.(II.19), the field equation (II.3) reads
$$(1/\sqrt{-{\cal G}})\partial_{\rho}\sqrt{-{\cal G}}\partial_{\rho}\phi_0
+(1/\sqrt{-{\cal G}})\partial_{\alpha}(\sqrt{-{\cal G}}{\cal G}^{\alpha \beta}
\partial_{\beta}\phi_1)-2\lambda[(3\phi_0^2-\eta^2)\phi_1 +3\phi_0\phi_1^2+
\phi^3_1]\ = \ 0\ .\eqno(II.26)$$
Replacing the above equation, to first order in $\phi_1$, in Eq.(II.23) we
obtain
$$L_2\ =\ -({1 \over 2} \sqrt{-{\cal G}}) [\phi_1 \partial_{\rho}\sqrt{-{\cal
G}}
\partial_{\rho}\phi_0) + \partial_{\alpha}(\phi_1 \sqrt{-{\cal G}}
{\cal G}^{\alpha \beta}\partial_{\beta}\phi_1]\ .\eqno(II.27)$$
We can now calculate the action
$$A\ =\ \int \sqrt{-{\cal G}}\  (L_0+L_1+L_2)d^4x\ ,\eqno(II.28)$$
where $\sqrt{-{\cal G}}$  is given by Eq.(II.18).
At this point we would like to remark that $\phi$ is a test field, since
Eq.(II.18) was derived using Einstein's equation in empty space. Replacing
Eq.(II.4) in the expression (II.21) we get that
$$L_0\ =\ -(\eta/\delta)^2 \cosh^{-4}(\rho/\delta)\ \eqno(II.29)$$
and, therefore, the first term in Eq.(II.28) for the action reads
$$\int\sqrt{-{\cal G}}L_0d^4x\ =\ -\mu\int\sqrt{-\gamma}d^3\tau+{1 \over 2}\int
\sqrt{-\gamma}^{(3)}Rd^3\tau \int\rho^2L_0(\rho)d\rho\ ,\eqno(II.30)$$
where the mass per unit area $\mu$ is given by Eq.(II.7). To get the second
term in the action, we use Eq.(II.25) and perform an integration by parts.
We get
$$\int \sqrt{-{\cal G}}L_1d^4x\ =\ \int\sqrt{-\gamma} K[\int \phi_1
\partial_{\rho}\phi_0 d\rho ]d^3\tau \ . \eqno(II.31)$$
Finally, the last term in the action becomes
$$\int \sqrt{-{\cal G}}L_2d^4x\ =\ -{1 \over 2}
\int\sqrt{-\gamma} K[\int \phi_1
\partial_{\rho}\phi_0 d\rho ]d^3\tau \ . \eqno(II.32)$$
where we have used Eq.(II.27). Combining the above results,
the action reads
$$A\ = \ -\mu \int\sqrt{-\gamma}d^3\tau+{1\over2} \int \sqrt{-\gamma} K [\int
\phi_1 \partial_{\rho}\phi_0 d\rho]d^3\tau + {1\over2}\int
\sqrt{-\gamma}^{(3)}Rd^3\tau \int\rho^2L_0(\rho)d\rho\ .\eqno(II.33)$$
The above expression for the action can be further simplified, under the
assumption of a domain wall having a small width $\delta $, arguing as
follows:
Let us first rewrite Eq.(II.33) as
$$A\ =\ -\mu \int\sqrt{-\gamma}d^3\tau
+{1\over2} \int \sqrt{-\gamma} K [\int
\phi_1 \partial_{\rho}\phi_0 d\rho]d^3\tau -\mu\alpha
\int \sqrt{-\gamma}^{(3)}Rd^3\tau \ , \eqno(II.34)$$
where $\alpha$ is given from
$$\alpha \mu \ = \  -{1\over2}\int \rho^2 L_0 d\rho \ = \ (\eta^2/2\delta^2)
\int_{- \infty}^{+ \infty}\rho^2 \cosh^{-4}
(\rho/\delta) d\rho \ \approx \ \eta^2\delta /3 $$
and thus, from Eq.(II.7)
$$\alpha\ \approx \ \delta^2/6\ .\eqno(II.35)$$
We can always expand $\phi_1$ in Taylor series as
$\phi_1=\Sigma_i \ \tilde\phi_i
\rho^i$. The term $i=0$ is absent ($\tilde\phi_0=0$) because  since the
domain wall is placed at the origin $\phi_1(\rho=0)=0$.
The general term contributes to the second integral of Eq.(II.34) as
$$\tilde\phi_i\int\rho^i\partial_{\rho} \phi_0 d\rho\
=\ \tilde\phi_i\eta\delta^i\int
({\rho\over\delta})^i\cosh^{-2}(\rho/\delta) d(\rho/\delta)\ ,\eqno(II.36)$$
where the integral is of order 1, for $i>2$ (for $i=1$ the integral vanishes
by parity). Thus the first term which contributes is for $i=2$, which is
of higher order in $\delta$ than the third integral in Eq.(II.34), so we
neglect it.
Thus the effective action simplifies to
$$A\ =\ -\mu \int\sqrt{-\gamma}d^3\tau - \mu\alpha
\int \sqrt{-\gamma}^{(3)}Rd^3\tau \ , \eqno(II.37)$$
where the first term is the usual Nambu action, while the second one represents
the curvature correction due to the small thickness of the domain wall.
To summarize, so far we have  obtained the effective action for a domain wall
with a small but non-zero thickness, embedded in an empty spacetime. At this
point we would like to mention that the form given by Eq.(II.37) for the
effective action with curvature corrections, had already appeared in the
literature. Our contribution was to do the analysis in a curved spacetime,
 give the explicit derivation of the expression for the effective action
and, mainly, find the sign and the order of magnitude of $\alpha$.

Let us now proceed with the calculation of the effective energy momentum
tensor. Our objective is to study the motion of a bubble wall described
by the action (II.37) in a curved spacetime. To do so, we will follow the
same analysis as in Ref. 8. We  start with the variation for the
effective action $A$, namely
$$\delta A \ =\ -(1/2) \mu\int\sqrt{-\gamma}\ [\gamma^{A B} - 2\alpha
^{(3)}G^{A B}]\delta\gamma_{A B}d^3\tau \ \eqno(II.38)$$
(where $^{(3)}G^{A B}$ stands for the induced Einstein tensor);
since the term with the divergence dropped out once we have integrated
over a closed surface.  As it was shown$^8$, for a purely geometrical
variation of the three dimensional worldsheet of the domain wall, the
variation of the induced metric reads
$$\delta_ {\rho}\gamma_{A B}\ = \ 2K_{A B} \ \delta\rho (\tau)\ .\eqno(II.39)$$
To get the equation of motion for the test bubble, we take the extremum
of the effective action A with respect to arbitrary variations in $\rho $.
The resulting equation reads
$$S^{A B}K_{A B}\ = \ 0\ , \eqno(II.40)$$
where
$$S^{A B}\ = \ -\mu \ [\gamma^{A B}- 2\alpha ^{(3)}G^{A B}]\ .\eqno(II.41)$$
Due to the invariance of the action under infinitessimal coordinate
transformations and under the assumption that the field satisfies the
equation of motion, one can show that the tensor $T^{\alpha \beta}$,
defined as
$$T^{\alpha \beta}\ =\  \left(\matrix{T^{A B}&0\cr
0&0}\right) , \eqno(II.42)$$
where
$$T^{A B}\ =\ S^{A B} \delta (\rho)\ ,\eqno(II.43)$$
is the conserved energy momentum tensor. Let us mention that the above
expression for the energy momentum tensor was also found by Letelier$^5$,
in his study of the motion of a test bubble in flat spacetime.
The quantity $S^{A B}$ is formally obtained in the usual way, but
using the induced metric
and is conserved, {\it i.e.,}
$$^{(3)}\nabla_A S^{A B}\ =\ 0\ ,\eqno(II.44)$$
where $^{(3)}\nabla_A $ denotes the three dimensional induced covariant
derivative.
\vskip 1 true cm
\centerline{III. Dynamics of spherical domain walls separating two geometries}
\vskip 1 true cm
In this section we will consider the gravitational effects of a spherically
symmetric domain wall with a small, non-zero, thickness separating two
spherically symmetric static geometries. We will then apply this formalism
in the particular case where these
geometries are de the Sitter and Schwarzschild metrics.
This case has been  studied earlier by Blau {\it et al}$^6$ and could probably
arise within the context of an inflationary scenario, in which
one may have chaotic cosmological initial conditions.
The difference between our study
and the detailed analysis presented in
Ref. 6 lies in the fact that we consider the
action of the domain wall taking into account the effect of curvature
corrections [see, Eq.(II.37)]. Suppose that the timelike hypersurface
$\Sigma $ separates the spacetime ${\cal M}$ into the manifolds ${\cal  M^+}$
and ${\cal M^-}$. As we have seen, the energy momentum tensor $T^{\alpha
\beta}$ [see, Eqs.(II.42), (II.43)] contains a $\delta$-singularity
without derivatives and therefore $\Sigma$  is a hypersurface of order one.
In other words, as we go through $\Sigma$, the metric of the spacetime
remains continuous, while the transverse derivatives are discontinuous.
Thus we can apply the formalism presented by Israel in Ref. 9.

Let $[A]=A_+-A_ - $ be the jump across the hypersurface $\Sigma$, and
$\tilde A=(1/2)(A_+ + A_ -)$  the average value of any discontinuous
quantity $A$. With this notation, the junction equations are
$$[K_{A B}]-\gamma_{A B}[K]\ =\ -8\pi G S_{A B}\ ;\eqno(III.1)$$
$$S_{A B} \tilde K ^{A B}\ = \ [t_{\alpha \beta}N^{\alpha}N^{\beta}]\ ;
\eqno(III.2)$$
$$^{(3)}\nabla_B S^B_{\ \ A}\ =\ -[t_{\alpha \beta}e^{\alpha}_{\ \ A}
N^{\beta}]\ ,\eqno(III.3)$$
where $t^{\pm}_{\alpha\beta}$  stand for the energy
momentum tensor of the continuous matter within the manifolds ${\cal M^{\pm}}$.
 In general, Eqs.(III.1)-(III.3) are not independent. Indeed, since
 the geometries ${\cal M^{\pm}}$
have known metrics $g^{\pm}_{\alpha \beta}$, which are
solutions of the Einstein equation associated with $t^{\pm}_{\alpha\beta}$,
one can easily check that Eqs.(III.2) and (III.3) will be
automatically satisfied once Eq.(III.1) is satisfied.
Hence we have to verify Eq.(III.1)
which is the Lanczos junction condition.
Moreover, due to  spherical symmetry, the off-diagonal
components of the extrinsic curvature vanish, while the angular ones are
related by
$$K_{\varphi\varphi}\ = \ \sin^2\vartheta K_{\vartheta \vartheta}\ .
\eqno(III.4)$$
So the dynamics of the domain wall will be completely determined by the
$(\tau\tau)$- and $(\vartheta\vartheta)$-components of the junction condition
(III.1). On the other hand, the spherical symmetry implies that the
surface energy tensor (II.41) cannot have $(\tau\vartheta)$- and
$(\tau\varphi)$-components and takes a form which resembles that
of a perfect fluid,
$$S^{A B}\ =\ (\sigma +p)u^A u^B + p\gamma^{A B}\ , \eqno(III.5)$$
where $\sigma, p, u=(1,0,0)$ denote the surface energy density,  surface
pressure and  3-velocity of a point in the domain wall respectively.
{}From Eqs.(II.41) and (III.5)
one gets
$$\sigma \ =\ \mu + 2\mu\alpha^{(3)}G^{A B}u_A u_B\ ,\eqno(III.6)$$
$$p\ =\ -\mu + (\mu\alpha/2)(2^{(3)}G^{A B}u_A u_B -^{(3)}R)\ .\eqno(III.7)$$
In the above expressions for $\sigma$ and $p$, the first term is the
usual Nambu term, while the second one is the result of the non-zero
thickness of the domain wall.

Let the manifold ${\cal M^{\pm}}$ have
the spherically symmetric static metric
$$ds^2_{\pm}\ =\ -f_{\pm}(r)dt^2_{\pm} + f^{-1}_{\pm}(r)dr^2
+r^2d\Omega^2\ ,\eqno(III.8)$$
where
$$d\Omega^2\ =\ d\vartheta ^2 +\sin^2\vartheta d\varphi^2\ .\eqno(III.9)$$
The timelike hypersurface $\Sigma$ generated by the spherical domain
wall is parametrized by
$$\tau^A\ =\ (\tau , \vartheta , \varphi)\ ,\eqno(III.10)$$
where $\tau$ denotes the time parameter as measured by an observer moving
with the domain wall. As a result of the matching condition on the induced
geometry, the metric on the hypersurface $\Sigma $ can be written as
$$ds^2_{\Sigma}\ =\ -d\tau^2 +R^2(\tau)d\Omega^2\ ,\eqno(III.11)$$
with $r=R(\tau)$.
 In the coordinates $(t,r,\vartheta,\varphi)$ the four components
 of the velocity $u$ introduced in Eq.(III.5) are
$u^{\alpha}_{\pm}\ = \ (\dot t_{\pm}, \dot R, 0, 0)$, where the
dot denotes derivation with respect to $\tau$. We introduce the notation
$$F_{\pm}\ \equiv \ \dot t_{\pm}f_{\pm}(R)\ =\ \epsilon^{\pm}_1 [f_{\pm}(R) +
\dot R^2]^{1/2}\ , \eqno(III.12)$$
where $\epsilon^{\pm}_1$ stands for the sign of $\dot t_{\pm}$ and the second
equality means that the square of the velocity $u^{\alpha}_{\pm}$ equals
$(-1)$.
The non-zero components of the extrinsic curvature are
$$K_{\tau}^{\pm \tau}\ =\ -\epsilon_2^{\pm}\
\dot F_{\pm}/\dot R \ ,\eqno(III.13)$$
$$K_{\vartheta}^{\pm \vartheta}\ =\ K_{\varphi}^{\pm \varphi}\ =
\  -\epsilon_2^{\pm}\ F_{\pm}/ R \ . \eqno(III.14)$$
Note that the sign indeterminacy $\epsilon_2^{\pm}$ depends on the expression
for the normal $N^{\alpha}_{\pm}$ to the hypersurface $\Sigma$, pointing
from ${\cal M^-}$ to ${\cal M^+}$. On the other hand, the non-zero components
of the induced Ricci tensor are:
$$^{(3)}R_{\tau}^{\tau}\ =\  2\ddot R/R\ ; \eqno(III.15)$$
$$^{(3)}R_{\vartheta}^{\vartheta}\ = \ ^{(3)}R_{\varphi}^{\varphi}\ =
\ (1/R^2)+(\dot R^2/R^2)+ \ddot R/R\ .\eqno(III.16)$$
The induced Ricci curvature scalar is
$$^{(3)}R\ =\ 2\ [2\ddot R/R + \dot R^2/R^2 + 1/R^2]\ \eqno(III.17)$$
and the non-zero components of the induced Einstein tensor read
$$^{(3)}G_{\tau}^{\tau}\ =\ -[\dot R^2/R^2 + 1/R^2]\ ; \eqno(III.18)$$
$$^{(3)}G_{\vartheta}^{\vartheta}\ =\  ^{(3)}G_{\varphi}^{\varphi}\ =
\ \ddot R/R\ .\eqno(III.19)$$
We can now rewrite the expressions [see, Eqs.(III.6) and (III.7)] for the
surface energy density $\sigma$ and the surface pressure $p$ as:
$$\sigma \ =\ \mu [1+2\alpha (\dot R^2/R^2+ 1/R^2)]\ ;\eqno(III.20)$$
$$p\ =\ -\mu[1+\alpha\ddot R/R]\ .\eqno(III.21)$$
Since both $\mu$ and $\alpha$ are positive, one can easily see that the
effect of the curvature corrections is to increase $\sigma$ with respect
to its value in the case of a Nambu domain wall. On the other hand,
the effect of the non-zero thickness of the domain wall on $p$,
depends on the sign of $\ddot R$. Stability conditions require $p$ to
be negative and this is satisfied
since $\alpha$ is small [see, Eq.(II.35)], provided we do not study
very small values of $R$.
 As we have already mentioned earlier on, the dynamics
of the domain wall will be completely determined once we solve the $(\tau
\tau)$- and $(\vartheta\vartheta)$-components of the Lanczos
junction condition (III.1).
The $(\vartheta\vartheta)$-component gives
$$-\epsilon_2^+ F_+ + \epsilon_2^- F_-\ =\  -4\pi\mu G [R+2\alpha(1/R
+\dot R^2/R)]\ ,\eqno(III.22)$$
and the $(\tau\tau)$-component becomes
$$-\epsilon_2^+\dot F_+ +\epsilon_2^-\dot F_-\ = \
-4\pi\mu G [\dot R+2\alpha\dot R(2\ddot R/R-1/R^2-\dot R^2/R^2)]\ ,
\eqno(III.23)$$
which is obviously the derivative of the $(\vartheta\vartheta)$-component.
Thus
the only equation we have to solve is Eq.(III.22).

Let us apply the
above analysis in the particular case of de Sitter and Schwarzschild geometries
joined with a thick spherical domain wall.  We   consider a
spherically symmetric region of false vacuum (de Sitter) which is separated
 from an infinite region of true vacuum (Schwarzchild) by a thick domain wall.
The Schwarzschild geometry is determined by
$$f_+\ =\ 1-(2GM/r)\ ,\eqno(III.24)$$
while
the de Sitter one by
$$f_-\ =\ 1-{\cal {X}}^2 r^2\ ,\eqno(III.25)$$
where
$${\cal {X}}^2\ = \ (8\pi/3)G\rho_0 \ ,\eqno(III.26)$$
with $\rho_0$ defined by the energy momentum tensor of ${\cal M^-}$,
{\it i.e.},
$$t^{-\alpha\beta}\ = -\rho_0 g_-^{\alpha\beta}\ .\eqno(III.27)$$
Doing some algebraic manipulations of Eq.(III.22) using Eqs.(III.12),
(III.24) and (III.25), where we keep only the linear terms in $\alpha$, since
$\alpha$ is small [see Eq.(II.31)], we obtain an equation of the form
$${\cal F}(R)\ = \ a(R)\dot R^2+b(R)\dot R^4\ , \eqno(III.28)$$
where
$${\cal F}(R)\ = \ (A-R\kappa)^2+(8GM/R)-[4+4R\kappa(A-R\kappa)\epsilon-
(16GM/R)\epsilon
+8\epsilon]\ ;\eqno(III.29a)$$
$$A(R)\ =\ (1/R\kappa)[(2GM/R-R^2{\cal X}^2]\ ;\eqno(III.29b)$$
$$\kappa\ = \ 4\pi\mu G\ ; \eqno(III.29c)$$
$$\epsilon\ =\  (2\alpha/R^2)\ ;\eqno(III.29d)$$
$$a(R)\ = \ 4+ 16 \epsilon +4R\kappa(A-R\kappa)\epsilon-(16GM/R)\epsilon\
;\eqno(III.29e)$$
$$b(R)\ = \ 8\epsilon\ .\eqno(III.29f)$$
As one can easily verify, Eq.(III.28) reduces to the equation of motion
of the domain wall found by Blau {\it et al}$^6$,  once we set $\alpha =0$.

Let us first find
 the positions $R$, where the velocity of the spherical domain wall vanishes.
These will be the solutions of the equation
$${\cal F}(R)\ =\ 0\ .\eqno(III.30)$$
To do so, we take $\lambda\sim 1$, the energy density of the false vacuum
$\rho_0$ to be of order $\eta^4$, the surface energy density $\mu$ to be
of order $\eta^3$ [see, Eqs.(II.5) and (II.6)] and we choose for the
Schwarzschild mass $M$ arbitrarily the value of the GUT scale, which is lower
than the critical mass $M_{crit} $, to be defined below.
For a domain wall formed during the spontaneous
symmetry breaking of GUT, $\eta\approx 10^{14}$GeV,
$\alpha\approx 10^{-28}(GeV)^{-2}$ and
Eq.(III.30) has two solutions [$R_1\approx
0.963\times10^{-15}  (GeV)^{-1}$ and $R_2\approx
 3.125\times10^{-14}  (GeV)^{-1}$],
almost identical to the case
of a zero thickness $(\alpha = 0)$ domain wall [$R^{\ast}_1\approx
 2.702\times10^{-15}  (GeV)^{-1}$ and $R^{\ast}_2\approx
 3.027\times10^{-14}  (GeV)^{-1}$].

At this point, we would like to remark that the equation
governing the dynamics of the bubble is a second order differential equation
[see, Eq.(III.23)]. As a consequence,
one has to draw the two dimensional phase space
diagram of Eq.(III.28), to get a qualitative analysis of
the domain wall dynamics. In Fig. 1 we show the phase
space diagram of that equation
for a Nambu domain wall and a domain wall with curvature corrections,
 respectively. For $M\ <\ M_{crit}$, which is the case under consideration,
 we found {\it bounded}, as well as {\it bounce} solutions.
Bounded solutions are the ones for which $R$ stars at zero, grows to the
maximum value $R^{\ast}_1 (R_1)$ and then returns to zero. On the other hand,
bounce solutions are
those for which $R$ approaches infinity in the asymptotic past, falls to
the minimum value given by $R^{\ast}_2 (R_2)$
and then approaches infinity in the
asymptotic future. This analysis holds for both a Nambu bubble, as
well as a bubble with curvature corrections, as one can easily verify
looking at Fig.1. We thus agree with the analysis performed by
 Blau {\it et al}$^6$ for a Nambu bubble, while we believe that our analysis
is rather simpler. The critical mass $M_{crit}$ is defined as the
Schwarzchild mass, for which the phase space diagram passes from the fixed
points of the two dimensional dynamical system describing the dynamics of the
bubble. The numerical value of  $M_{crit}$, for a Nambu domain wall ($\alpha
=0$),
 is found by demanding
${\cal F}(R)\ =\ 0$ and $d{\cal F}(R)/dR\ =\ 0$ and it is $M_{crit}\ \sim\
10^{28}GeV$. The numerical value of the critical mass for a domain wall with
curvature corrections is of the same order of magnitude. As one can see from
Fig.1, the effect of the curvature corrections on the bubble dynamics is to
decrease $R^{\ast}_1$ and increase $R^{\ast}_2$. This could be interpreted as
a global dragging effect caused by the increase of the surface energy density
and the modification of the tension.

\vskip 1 true cm
\centerline {Acknowledgments}
\vskip 0.6 true cm
We would like to thank Brandon Carter and Hector Giacomini for
stimulating discussions.
\vskip 1 true cm
\centerline { References}
\vskip 0.6 true cm
\noindent $^{\diamond}$Electronic address: barrabes@univ-tours.fr
\vskip 0.2 true cm
\noindent $^{\triangleleft}$Electronic address: boisseau@univ-tours.fr
\vskip 0.2 true cm
\noindent $^{\star}$Electronic address: sakellar@univ-tours.fr
\vskip 0.2 true cm
\noindent $^{1}$S. Coleman, Phys. Rev. D {\bf 15}, 2929 (1977).
\vskip 0.2 true cm
\noindent $^{2}$R. Gregory, Phys. Rev. D {\bf 43}, 520 (1991); R. Gregory,
D. Haws and D. Garfinkle, Phys. Rev. D {\bf 42}, 343 (1990).
\vskip 0.2 true cm
\noindent $^{3}$V. Silveira and M. D. Maia, Phys. Lett. A {\bf 174}, 280
(1993).
\vskip 0.2 true cm
\noindent $^{4}$A. L. Larsen, "Comment on thickness-corrections to Nambu-
wall" (preprint) (June 1993).
\vskip 0.2 true cm
\noindent $^{5}$P. S. Letelier, Phys. Rev. D {\bf 41}, 1333 (1990).
\vskip 0.2 true cm
\noindent $^{6}$S. K. Blau, E. I. Guendelman and A. H. Guth, Phys. Rev. D
{\bf 35}, 1747 (1987).
\vskip 0.2 true cm
\noindent $^{7}$Y. B. Zel'dovich, I. Y. Kobzarev and L. B. Okum, Sov. Phys.
JETP {\bf 40}, 1 (1975).
\vskip 0.2 true cm
\noindent $^{8}$ B. Boisseau and P. S.  Letelier, Phys. Rev. D {\bf 46},
 1721 (1992).
\vskip 0.2 truecm
\noindent $^{9}$ W. Israel, Nuovo Cimento {\bf 44}, 1 (1966).
\end